\documentclass[twocolumn,prl,superscriptaddress,showpacs]{revtex4}

\usepackage{euscript,amssymb,amsmath}
\newcommand{\diag}{\mbox{\rm diag}\,}

\newcommand{\be}[1]{\begin{equation}\label{#1}}
\newcommand{\ee}{\end{equation}}
\newcommand{\ba}[1]{\begin{eqnarray}\label{#1}}
\newcommand{\ea}{\end{eqnarray}}
\newcommand{\rf}[1]{(\ref{#1})}
\newcommand{\nn}{\nonumber}
\newcommand{\dd}{\dagger}

\def\RR{\mathbb{R}}

\def\CC{\mathbb{C}}

\def\a{\alpha}
\def\b{\beta}
\def\d{\delta}

\def\sg{\sigma}
\def\e{\varepsilon}

\def\om{\omega}

\newcommand{\cH}{\mathcal{H}}

\newcommand{\cM}{\mathcal{M}}
\newcommand{\cN}{\mathcal{N}}
\newcommand{\cP}{\mathcal{P}}
\newcommand{\cT}{\mathcal{T}}

\newcommand{\bfphi}{\boldsymbol{\phi}}
\newcommand{\bfpsi}{\boldsymbol{\psi}}
\newcommand{\bfPhi}{\boldsymbol{\Phi}}

\newcommand{\bE}{\mathbf{E}}

\newcommand{\bH}{\mathbf{H}}
\newcommand{\bI}{\mathbf{I}}

\newcommand{\bP}{\mathbf{P}}

\newcommand{\bU}{\mathbf{U}}
\newcommand{\bV}{\mathbf{V}}

\newcommand{\ra}{\rangle}
\newcommand{\la}{\langle}

\begin{document}

\title{The Naimark dilated $\cP\cT-$symmetric bra\-chi\-sto\-chrone}

\author{Uwe G\"unther}
\email{u.guenther@fzd.de}
\affiliation{Research Center
Dresden-Rossendorf, POB 510119, D-01314 Dresden, Germany}
\author{Boris F. Samsonov}
\email{samsonov@phys.tsu.ru} \affiliation{Tomsk State University, 36
Lenin Avenue, 634050 Tomsk, Russia}

\date{July 23, 2008}

\begin{abstract}The quantum mechanical brachistochrone system with $\cP\cT-$symmetric
Hamiltonian is Naimark dilated and reinterpreted as subsystem of a
Hermitian system in a higher-dimensional Hilbert space. This opens a
way to a direct experimental implementation of the recently
hypothesized $\cP\cT-$symmetric ultra-fast brachistochrone regime of
[C. M. Bender et al, Phys. Rev. Lett. {\bf 98}, 040403 (2007)] in an
entangled two-spin system.
\end{abstract}
 \maketitle

{\em Introduction}\quad The quantum brachistochrone problem consists
in finding a Hamiltonian $H$ which evolves a given initial state
$|\psi_I\ra$ into a given final state $|\psi_F\ra$ in a minimal time
$\tau$. Considering this problem for quantum mechanics with
$\cP\cT-$symmetric Hamiltonians (PTQM) Bender, Brody, Jones and
Meister (BBJM) found the surprising result \cite{cmb-brach} that the
minimal evolution (passage) time $\tau_{\cP\cT}$ was less than the
minimal time $\tau_h$ required for the evolution induced by a
Hermitian Hamiltonian \cite{brody-1,jap-brach}.  It could be made
even arbitrary small $\tau_{\cP\cT}\to 0$ in a strongly
non-Hermitian regime \cite{cmb-brach,GRS-jpa-ep}. If this effect of
a `faster than Hermitian' evolution \cite{cmb-brach} were
experimentally realizable it would open a way to ultra-fast quantum
computing processes \cite{Nielsen}. A problem still unsolved in
\cite{cmb-brach} concerned the switching mechanism between the
$\cP\cT-$symmetric brachistochrone system and a conventional (von
Neumann) quantum system necessary for an experimental implementation
of the suggested ultra-fast quantum process.

As shown by Mostafazadeh \cite{most-brach}, an equivalence mapping
\cite{ali-herm-1} between PTQM in the sector of unbroken
$\cP\cT-$symmetry and conventional quantum mechanics (CQM) leaves
the passage time of a brachistochrone invariant
$\tau_h=\tau_{\cP\cT}$. This implies that a vanishing passage time
$\tau_{\cP\cT}\to 0$ in the $\cP\cT-$symmetric system is necessarily
connected with a vanishing distance between initial and finite state
in the equivalent Hermitian system --- an effect geometrically
analyzed in \cite{GS-geom-brach}. In case of the Hermitian
equivalent of the BBJM brachistochrone, initial and final states
will nearly coincide (coincidence problem) so that the
brachistochrone effect in such an interpretation would loose any
physical relevance.

In this Letter, we propose a realization of the BBJM brachistochrone
\cite{cmb-brach} which resolves the switching problem between PTQM
and CQM regimes \cite{cmb-brach}, avoids the coincidence problem
\cite{most-brach,GS-geom-brach} and which can be considered as
starting point for a direct experimental implementation. The key
idea consists in a reinterpretation of the BBJM brachistochrone as
$\cP\cT-$symmetric subsystem of a larger CQM system living in a
higher-dimensional Hilbert space. For this purpose we use a Naimark
dilatation (extension) technique \cite{Holevo} as it is widely used
in quantum information theory \cite{Nielsen}. We will demonstrate
that the resulting large system will have the structure of an
entangled two-spin (two-qubit) system so that an experimental
realization of the BBJM brachistochrone effect should be feasible,
e.g., in a suitably designed system of entangled polarized photons
\cite{zeilinger-unitary-experiment}.

Technically, the construction of the large Hermitian system will be
accomplished by a three-step procedure: (i) by building a suitable
positive operator valued measure (POVM) \cite{Nielsen,Holevo,POVM}
over the nonorthogonal eigenstates of the $\cP\cT-$symmetric
Hamiltonian $H$ and its adjoint $H^\dd$, (ii) by Naimark dilating
(extending) \cite{Holevo} this POVM into an orthogonal projector set
in the higher dimensional Hilbert space and (iii) by constructing
from it a corresponding Hermitian Hamiltonian $\bH=\bH^\dd$ and a
unitary evolution operator $\bU(t)=e^{-it\bH }$.

{\em BBJM brachistochrone}\quad The BBJM brachistochrone
\cite{cmb-brach} that we are going to Hermitianly dilate (extend)
describes the evolution from an initial state $|\psi_I\ra$ to a
final state $|\psi_F\ra$ governed by a $\cP\cT-$symmetric
Hamiltonian $H$ whose structure is chosen in such a way that the
time $\tau$ required for the evolution becomes minimal. As shown in
\cite{brody-1} such a minimal-passage-time solution follows a
minimal geodesic in projective Hilbert space and it is therefore
located in the two-dimensional subspace $\cH_2=\CC^2$ spanned by
$|\psi_I\ra$ and $|\psi_F\ra$. In this $\cH_2$ the
$\cP\cT-$symmetric Hamiltonian $H$ can be chosen as
\cite{cmb-brach,GRS-jpa-ep}
\be{1}
H=E_0I_2+s\left(%
\begin{array}{cc}
  i\sin(\a) & 1 \\
  1 & -i\sin(\a) \\
\end{array}%
\right),\ E_0,s\in\RR,
\ee where $\cP=\sg_x$ denotes the parity operator, $\cT$ is the antilinear operator of
time reflection and complex conjugation \cite{cmb-rev}, $E_0$
denotes an irrelevant offset energy and $s$ a general scaling factor
of the matrix. (As usual, $\sg_x$, $\sg_y$ and $\sg_z$ are Pauli
matrices.) The angle $\a\in (-\pi/2,\pi/2)$ characterizes the
non-Hermiticity of the Hamiltonian:  $H(\a=0)$ is Hermitian, whereas
in the limit $\a\to \pm \pi/2$ the Hamiltonian $H$ becomes strongly
non-Hermitian and similar to a Jordan block, i.e. its eigenvectors
\ba{2}
|E_+(\a)\ra&=&\frac{e^{i\a/2}}{\sqrt{2\cos(\a)}}\left(
          \begin{array}{c}
            1 \\
            e^{-i\a} \\
          \end{array}
        \right)\nn\\ |E_-(\a)\ra&=&\frac{ie^{-i\a/2}}{\sqrt{2\cos(\a)}}\left(
          \begin{array}{c}
            1 \\
            -e^{i\a} \\
          \end{array}
        \right)
\ea
and eigenvalues $E_\pm=E_0\pm s\cos(\a)=:E_0\pm\om_0/2$ coalesce for
fixed $|s|<\infty$  \cite{GRS-jpa-ep}. The Hamiltonian is restricted
to purely real eigenvalues, i.e. the parameter sector of exact
$\cP\cT-$symmetry \cite{cmb-rev}. The operator $U(t)=e^{-itH}$ of
the non-unitary evolution induced by $H$ has the explicit form
\be{3}
U(t)=\frac{e^{-iE_0t}}{\cos(\a)}\left(
                                                                 \begin{array}{cc}
                                                                   \cos(y-\a) & -i\sin(y) \\
                                                                   -i\sin(y)& \cos(y+\a) \\
                                                                 \end{array}
                                                               \right)
\ee
with $y:=\om_0 t/2$ (we set $\hbar=1$). In the
BBJM-bra\-chis\-to\-chrone setup \cite{cmb-brach} this $U(t)$ is
used to evolve an initial state $|\psi_I\ra=(1,0)^T$ into a final
state $|\psi_F\ra=\mu_F(0,1)^T$, $\mu_F:=- ie^{-iE_0 \tau}$.  The
time $\tau$ required for this evolution follows from the condition
$y=\a +\pi/2$ as $\tau=\frac{\a+ \pi/2}{s\cos(\a)}$ and tends for
\be{4}
\a=\e- \pi/2,\quad \e\to +0
\ee
and fixed $s\cos(\a)=\om_0/2$ to zero: $\tau\to 0$. In this way the
evolution from $|\psi_I\ra$ to the orthogonal $|\psi_F\ra$ induced
by the $\cP\cT-$symmetric Hamiltonian $H$ with eigenstates of fixed
energy difference $E_+-E_-=\om_0$ appears faster than an evolution
between these states induced by any Hermitian Hamiltonian with the
same energy difference $\om_0$ between its eigenstates. This is due
to the fact that the evolution time $\tau$ between orthogonal states
in Hermitian system has to be larger than the Anandan-Aharonov lower
bound $\tau\ge \tau_h=\pi/\om_0$
\cite{brody-1,anandan-aharonov-prl-1990}.

Before we embed the BBJM-brachistochrone into a larger Hermitian
model we briefly collect the required setup information. The
eigenvectors \rf{2} of $H$ are normalized with regard to the
$\cP\cT$ inner product $(u,v)=\cP\cT u\cdot v$ as $
(E_\pm,E_\pm)=\pm 1,\ (E_\pm,E_\mp)=0$  \cite{cmb-rev} and for $\a
\neq 0$ they are nonorthogonal with regard to the standard inner
product in the Hilbert space $\cH_2=\CC^2$: $\la
E_\pm|E_\mp\ra\neq 0$. We supplement them via $H^\dd(\a)=H(-\a)$
with the eigenvectors $|E_+(-\a)\ra$, $|E_-(-\a)\ra$ of the
adjoint operator $H^\dd$ and arrange them as columns in the
matrices
\be{5}\Psi:=\left[\,|E_+(\a)\ra,|E_-(\a)\ra\right],\
\Xi:=\left[\,|E_+(-\a)\ra,|E_-(-\a)\ra\right].
\ee
With $\tilde E:=\diag(E_+,E_-)$ the eigenvalue problems for $H$
and $H^\dd$ take then the compact matrix form
\be{6}
H\Psi=\Psi \tilde E,\qquad H^\dd\Xi=\Xi \tilde E.
\ee
Apart from the bi-orthonormality relation $\Xi^\dd\Psi=I_2$, it
holds $ \Psi\Psi^\dd H^\dd=H\Psi\Psi^\dd$,\ $\Xi\Xi^\dd H=H^\dd
\Xi\Xi^\dd$ so that one identifies
$\left(\Psi\Psi^\dd\right)^{-1}=\Xi\Xi^\dd=\eta$ as metric operator
in the pseudo-Hermiticity condition $\eta H=H^\dd \eta$
\cite{ali-herm-1}. Additionally to its obvious Hermiticity
$\eta=\eta^\dd$ the metric can be suitably scaled to be an element
of the hyperbolic (``boost'') sector of the complex orthogonal group
$SO(2,\CC)$ \cite{GS-geom-brach}
\ba{8}
\eta&=&\frac1{\cos(\a)}\left(%
\begin{array}{cc}
  1 & -i\sin(\a) \\
  i\sin(\a) & 1 \\
\end{array}%
\right)\nn\\
&=&\left(%
\begin{array}{cc}
  \cosh(\b) & -i\sinh(\b) \\
  i\sinh(\b) & \cosh(\b) \\
\end{array}%
\right)=e^{\b\sigma_y}
\ea
with parameter identification $\sin(\a)=:\tanh(\b)$ and
$\cos(\a)=1/[\cosh(\b)]$. As final ingredient we fix the notation
for the one-to-one similarity mapping between the $\cP\cT-$symmetric
Hamiltonian $H$ and its isospectral Hermitian counterpart $h=h^\dd$
\cite{ali-herm-1}, $H=\rho^{-1} h\rho$,\ $H^\dd=\rho h\rho^{-1}$,\
$\rho^2=\eta $, as well as for the unitary eigenvector matrix $\Phi$
\ba{10}
h\Phi=\Phi\tilde E,\qquad \Phi^\dd=\Phi^{-1}.
\ea
The eigenvectors of $H$ and $H^\dd$ can be regarded  as
$\a\rightleftharpoons -\a$, i.e. $\b\rightleftharpoons -\b$, mirror
symmetrically distorted versions of the eigenvectors of the
Hamiltonian $h$
\be{11}
\Psi=\rho^{-1}\Phi, \qquad \Xi=\rho \Phi,\qquad
\rho^{-1}(\b)=\rho(-\b).
\ee

The orthogonal initial and final vectors $|\psi_I\ra$, $|\psi_F\ra$
in the BBJM-brachistochrone model on their turn can be considered as
eigenstates of a Hermitian spin operator $S_z=\sg_z$ (a von-Neumann
observable with orthogonal projector decomposition), whereas the
$\cP\cT-$symmetric (non-Hermitian) Hamiltonian $H$ has nonorthogonal
eigenvectors $|E_\pm\ra$ and is not a von-Neumann observable. Under
the one-to-one equivalence mapping \cite{ali-herm-1} from $H$ to the
Hermitian Hamiltonian $h$ the spin operator $S_z$ maps into a
non-Hermitian operator $s_z=\rho S_z\rho^{-1}\neq s_z^\dd$. Hence,
the BBJM-brachistochrone system in both representations $(H,S_z)$
and $(h,s_z)$ contains operators which are not von-Neumann
observables and therefore the system cannot be considered as
fundamental.

{\em Naimark dilation}\quad In order to give the BBJM system with
evolution $\psi(t)=U(t)\psi_I$ a meaning in CQM we embed it into a
larger purely Hermitian system
\ba{12}
\hat\bfpsi(t)=\bU(t)\hat\bfpsi_I,\qquad \hat\bfpsi(t)=\left(%
\begin{array}{c}
  \psi(t) \\
  \chi(t) \\
\end{array}%
\right)
\ea
with unitary evolution operator
$\bU(t)=\left[\bU^\dd(t)\right]^{-1}$ and additional ancilla wave
function component $\chi(t)$. For this purpose we construct an
auxiliary POVM \cite{Nielsen,Holevo,POVM} $\sum_{k=1}^4 A_k=I_2$
over the nonorthogonal eigenvectors \rf{2}, \rf{5} of $H$ and its
adjoint $H^\dd$ with rank-one operators $A_{1,2}=f^2|E_\pm(\a)\ra\la
E_\pm(\a)|$ and $A_{3,4}=f^2|E_\pm(-\a)\ra\la E_\pm(-\a)|$. For
symmetry reasons all $A_k$ are scaled with the same normalization
factor
\be{12b}
f:=\sqrt{\frac{\cos(\a)}2}=\frac 1{\sqrt{2\cosh(\b)}}\,.
\ee
Next we note that any rank-one-type POVM built from a general vector
set $\{x_k\}_{k=1}^{n>N}$, $x_k\in \CC^N$ reads in matrix form $
\sum_{k=1}^{n>N}x_k x_k^\dd=MM^\dd=I_N$,\ $ x_k\in\CC^N$,\ $
M:=[x_1,x_2,\ldots,x_n]\in \CC^{N\times n}$,  where $M$ describes a
partial matrix isometry $M:\ \CC^n\to \CC^N$ with $n>N$. The lifting
of this isometry to a unitary mapping $\bV: \CC^n\to \CC^n$ is known
as Naimark dilation (extension) \cite{Holevo} and can be
accomplished by a suitable embedding $
M\hookrightarrow \bV=\left[%
\begin{array}{c}
  M \\
  L \\
\end{array}%
\right]$, \ $L\in\CC^{(n-N)\times n}$ with additional unitarity
constraint
\be{14}
\bV\bV^\dd=\bV^\dd \bV=\bI_n.
\ee
For our $\CC^{2\times 4}$ setup the matrix $M$ can be read off from
eqs. \rf{5} and \rf{11} as $ M=f[\Psi,\Xi]=f[\rho^{-1},\rho]\bfPhi
$,
where $\bfPhi:=\left(%
\begin{array}{cc}
  \Phi & 0 \\
  0 & \Phi \\
\end{array}%
\right)=I_2\otimes \Phi.$ The natural ansatz
$
\bV=f\left(%
\begin{array}{cc}
  \rho^{-1} & \rho \\
  X & Y \\
\end{array}%
\right)\bfPhi$ together with the auxiliary condition $
MM^\dd=f^2[\rho^2+\rho^{-2}]=f^2[\eta+\eta^{-1}]=I_2$ (following
from \rf{8}, \rf{10} and \rf{12b}) and the constraint \rf{14} fix
the nonsingular matrices $X$, $Y$ up to an irrelevant unitary
rotation as $X=\rho$, $Y=-\rho^{-1}$ so that
\ba{19}
\bV&:=&f\left[\sg_z\otimes \rho^{-1}+\sg_x\otimes
\rho\right](I_2\otimes\Phi)\nn\\
&=&f\left[\sg_z\otimes \Psi+\sg_x\otimes \Xi\right]\,.
\ea
The columns of $\bV=[v_1,v_2,v_3,v_4]$ are formed by four orthogonal
vectors  $v_k\in\cH_4=\CC^4$ which provide the desired Naimark
dilations of the nonorthogonal eigenvectors of $H$ and its adjoint
$H^\dd$ living in $\cH_2=\CC^2$. Additionally, they yield the
embedding $A_k\hookrightarrow P_k=v_kv_k^\dd$ of the $\cH_2-$POVM
into the ortho-projector set $P_jP_k=\d_{jk}P_k$,
$\sum_{k=1}^4P_k=I_4$ in $\cH_4$ \cite{Holevo}.

We start the construction of a selfconsistent CQM in $\cH_4$ by
requiring that the original eigenvalue problems for $H$  and $H^\dd$
are recovered when the model is restricted to the first two rows of
$\bV$. From relations \rf{6} and an ansatz
$f[H\Psi,H^\dd\Xi]=f[\Psi\tilde E,\Xi\tilde E]=M\bE$ the eigenvalue
matrix $\bE$ for the dilated problem can be read off as $\bE:=\left(
       \begin{array}{cc}
         \tilde E & 0 \\
         0 & \tilde E \\
       \end{array}
     \right)=I_2\otimes \tilde E$.
This means that the corresponding dilated Hamiltonian $\bH$ will
have the two eigenvalues $E_\pm$ of $H$ and its isospectral adjoint
$H^\dd$ as double degenerate eigenvalues. The Hamiltonian $\bH$
itself can be constructed from eq. \rf{10} and $\bH\bV=\bV\bE$ as $
\bH=\bV\bE\bV^\dd$ so that
\ba{22}
\bH&=&f^2\left[I_2\otimes (H\eta^{-1}+\eta H)+i\sg_y\otimes
(H-H^\dd)\right]\nn\\
  &=:&I_2\otimes\Lambda +i\sg_y\otimes\Omega \nn
\ea
where \ba{22e} \Lambda&:=&f^2(H\eta^{-1}+\eta H)=E_0
I_2+\frac{\om_0}2\cos(\a)\sg_x\nn\\
\Omega&:=&f^2(H-H^\dd)=i\frac{\om_0}2\sin(\a) \sg_z.\nn
\ea
This $\bH$ is Hermitian by construction. In the Hermitian limit of
the original $\cP\cT-$symmetric Hamiltonian $H$, i.e. for
$\a=\b=0$, it holds $\eta=I_2$ and $\bH$ reduces to
$\bH=I_2\otimes h$ --- a trivially doubled $h$. In contrast to the
PTQM Hamiltonian $H$ its dilation $\bH$ remains well defined also
in the strongly non-Hermitian vanishing-passage-time regime \rf{4}
where the matrix components of $H$ diverge for fixed $\om_0$ as
$s\to\infty$. This regularization effect is due to the
normalization factor $f^2$ induced in $\bH$ via the auxiliary POVM
construction.

The $\bH-$induced unitary evolution in $\cH_4$ is governed by the
operator  $\bU(t)=e^{-it\bH }=\bV e^{-i\bE t} \bV^\dd$ which via
\rf{3}, \rf{8}, \rf{10}, \rf{12b},
$U(t)=e^{-itH}=\rho^{-1}e^{-ith}\rho$ and $y=\om_0 t/2$ can be
represented as
\ba{24}
\bU(t)&=&f^2\left\{I_2\otimes \left[U(t)\eta^{-1}+\eta
U(t)\right]\right.+\nn\\
&&\left. +i\sg_y\otimes \left[U(t)-\eta U(t)
\eta^{-1}\right]\right\}\nn\\
 &=&\left(I_2\otimes F+i\sg_y\otimes G\right)=\left(
                                                                        \begin{array}{cc}
                                                                          F & G \\
                                                                          -G & F \\
                                                                        \end{array}
                                                                      \right)
\nn\\
F&:=& e^{-iE_0 t}\left[I_2 \cos(y)-i\sg_x
\sin(y)\cos(\a)\right]\nn\\
G&:=& e^{-iE_0 t}\left[\sin(y)\sin(\a)\sg_z\right].
\ea
Physically $\bU(t)$ describes the time evolution of the coupled
brachistochrone-ancilla system \rf{12} in a Hilbert space
$\cH_4=\cH_2\oplus\tilde \cH_2$ with $\psi(t)\in\cH_2$ and
$\chi(t)\in\tilde\cH_2$. In order to exactly reproduce the
$\cH_2-$evolution \rf{3} of the BBJM-brachistochrone subsystem
\be{24a}
\psi(t)=U(t)\psi_I=\frac{e^{-iE_0 t}}{\cos(\a)}\left(
                                                         \begin{array}{c}
                                                           \cos(y-\a) \\
                                                           -i\sin(y) \\
                                                         \end{array}
                                                       \right)
\ee
the initial vector $\chi_I\in \tilde \cH_2$ of the ancilla subsystem
should be chosen appropriately. To obtain $\chi_I$ we represent
$\hat\bfpsi(t)\in\cH_4$ as
\be{25}   \hat\bfpsi(t)=\left(
       \begin{array}{c}
         \psi(t) \\
         \chi(t) \\
       \end{array}
     \right)=e_+\otimes \psi(t)+e_-\otimes \chi(t)
\ee
with $e_+:=(1,0)^T$ and $e_-:=(0,1)^T$, define $P_\pm:=e_\pm\otimes
e_\pm^\dd$ and introduce the projectors $\bP_\pm=P_\pm\otimes I_2$
on the brachistochrone ($\bP_+$) and the ancilla $(\bP_-)$ subspace.
The identification rule \rf{12} takes then the form $\bP_+
\hat\bfpsi(t)=\bP_+ \bU(t)\hat\bfpsi_I =e_+\otimes
\psi(t)=e_+\otimes U(t)\psi_I $. After taking the time derivative
one finds from
\be{28}
\bP_+ \bH \hat\bfpsi(t) \equiv e_+\otimes [\Lambda \psi(t)+\Omega
\chi(t)]=e_+\otimes H \psi(t)
\ee
a synchronization link\footnote{Comparison with the relations
\rf{11} between the $\cP\cT-$symmetric subsystem and its adjoint,
i.e. with $\Xi=\eta\Psi$,  shows that the ancilla $\chi(t)$ can be
interpreted as adjoint to BBJM brachistochrone.}  between ancilla
and brachistochrone evolution $
\chi(t)=\Omega^{-1}(H-\Lambda)\psi(t)=\eta\psi(t) $ as well as the
explicit ancilla evolution
$\chi(t)=\frac{e^{-iE_0 t}}{\cos(\a)}\left(%
\begin{array}{c}
  \cos(y) \\
  -i\sin(y-\a) \\
\end{array}%
\right)$. Initial and final ancilla component take then the form
$
\chi_I=\eta\psi_I=\frac1{\cos(\a)}\left(%
\begin{array}{c}
  1 \\
  i\sin(\a) \\
\end{array}%
\right)$, $
\chi_F=\eta\psi_F=\frac{-\mu}{\cos(\a)}\left(%
\begin{array}{c}
  \sin(\a) \\
  i \\
\end{array}%
\right) $ with $\mu=e^{-iE_0 \tau}$ an irrelevant phase factor and
$\psi_I=(1,0)^T$, $\psi_F=\mu (0,1)^T$.

{\em Discussion}\quad For a BBJM-brachistochrone in the vanishing
passage time regime \rf{4} the ancilla vectors $\chi_I$ and $\chi_F$
become collinear and their common denominator $\cos(\a)\approx \e$
makes them highly dominant compared to $\psi_{I,F}$. This
$\chi-$dominance remains preserved for the normalized state vector
$\hat\bfphi:=g\hat\bfpsi$, $\la\bfphi|\bfphi\ra=1$, with
$g:=\cos(\a)/\sqrt 2$ and leads to a very small brachistochrone
component $|g\psi(t)|^2\approx \e^2/2$ compared to the ancilla
component $|g\chi(t)|^2\approx 1-\e^2/2$. As result the geodesic
distance between the initial and final states in $\cH_4$ becomes
small $ \d_4=2\arccos(|\la \hat\bfphi_I|\hat\bfphi_F\ra|)\approx
2\e$.  This means that the original geodesic distance
$\d_2=2\arccos(|\la \psi_I|\psi_F\ra|)=\pi$ between the initial and
final states in the brachistochrone subsystem is strongly contracted
by embedding the latter into the larger Hermitian $\cH_4-$system.
Geometrically, this follows from the fact that the geodesic distance
is given by the angle spanned by the corresponding vectors on the
Bloch sphere \cite{brody-1} and its generalization to higher
dimensions. In the $\cH_2-$subsystem the vectors $\psi_I$ and
$\psi_F$ are antipodal and span an angle $\d_2=\pi$. Adding a much
longer vector $\chi$ orthogonal to $\psi_I$ and $\psi_F$ makes the
resulting  $\hat\bfphi_I\approx (\psi_I,\chi)^T$ and
$\hat\bfphi_F\approx (\psi_F,\chi)^T$ almost collinear $\d_4\to 0$
in $\cH_4$. In this way the dilated model reconciles the
Aharonov-Anandan lower bound
\cite{brody-1,anandan-aharonov-prl-1990} on minimal passage times in
Hermitian systems with the vanishing passage time effect of the
BBJM-brachistochrone \cite{cmb-brach} for orthogonal states in the
subsystem. The embedding of the BBJM-system into a
higher-dimensional Hilbert space can be regarded as a strengthening
of the wormhole analogy drawn in \cite{cmb-brach} for the shortening
of the passage time $\tau$.  A wormhole connection of two distant
points on a given lower dimensional manifold $\cM$ can be best
visualized by embedding $\cM$ into a higher dimensional surrounding
$\cN\supset\cM$ so that the corresponding short distance in $\cN$
becomes obvious \cite{thorne}.

The representation \rf{25} indicates on a natural interpretation of
the obtained Hermitian system $\hat\bfphi(t)=e^{-it\bH}\hat
\bfphi_I$ as system of two entangled spin $1/2$ particles, i.e. as a
two-qubit system \cite{Nielsen}, with $\Sigma_1=\sg_z\otimes I_2$
and $\Sigma_2=I_2\otimes \sg_z$ as spin operators of the two
spin-subsystems. In order to observe the BBJM-brachistochrone effect
of the subsystem one has to prepare an initial entangled state
$\hat\bfphi_I=e_+\otimes \psi_I+e_-\otimes\chi_I$, to switch on the
interaction Hamiltonian $\bH$ during the passage time $\tau$
(assumed as smaller than the lower passage time bound
$\tau_h=\pi/\om_0$) and to evolve $\hat\bfphi_I$ into the final
state $\hat\bfphi_F=e^{-i\tau\bH}\hat\bfphi_I$. This final state has
to be analyzed in a two-step measurement. In a first (instantaneous)
$\Sigma_1-$measurement one selects (filters out) the up-component
$e_+$ of the first spin. This results in a state
$\bP_+\hat\bfphi_F/\la\bP_+\hat\bfphi_F|\bP_+\hat\bfphi_F\ra
^{1/2}=e_+\otimes \psi_F/\la\psi_F|\psi_F\ra^{1/2}$ and separates
the brachistochrone component from the ancilla component (connected
with the down-component $e_-$ of the first spin). In a subsequent
$\Sigma_2-$measurement, one analyzes the spin-up and spin-down
states of the brachistochrone component $
\psi_F/\la\psi_F|\psi_F\ra^{1/2}$ to recover the spin-flip effect
from $\psi_I=(1,0)^T$ to $\psi_F=\mu_F(0,1)^T$.

A direct experimental test should be feasible with a suitably
designed system of entangled photons passing an appropriately chosen
arrangement of beam splitters, phase shifters, and mirrors as
implementation of the unitary operator $\bU(\tau)=e^{-i\tau\bH}$
\cite{zeilinger-unitary-experiment}.

{\em Conclusions}\quad We have demonstrated that the quantum
brachistochrone for a system with $\cP\cT-$symmetric Hamiltonian can
be realized as subsystem of a larger Hermitian system living in a
higher-dimensional Hilbert space. The Hermitian system (constructed
by Naimark dilating an auxiliary positive operator valued measure)
has the structure of an entangled two-quibit system. This opens a
way to direct experimental tests on the recently hypothesized
`faster than Hermitian' evolution in $\cP\cT-$symmetric quantum
systems.

The work of UG has been supported by DFG within the  Collaborative
Research Center SFB 609. BFS is partially supported by the grants
RFBR-06-02-16719, SS-871.2008.2 (Russia) and 4-7531.50-04-844-07/5
(Saxon Ministry of Science).


\end{document}